\documentclass{iaus}
\usepackage{graphicx}

\title[X-ray Emission from O Stars] 
{X-ray Emission from O Stars}

\author[David H. Cohen]   
{David H. Cohen$^1$}

\affiliation{$^1$Swarthmore College, Department of Physics and
  Astronomy, 500 College Ave., Swarthmore, Pennsylvania 19081, USA}

\pubyear{2008}
\volume{250}  
\jname{Massive Stars as Cosmic Engines}
\editors{F. Bresolin, P.A. Crowther \& J. Puls, eds.}
\begin{document}

\maketitle

\begin{abstract}

  Young O stars are strong, hard, and variable X-ray sources,
  properties which strongly affect their circumstellar and galactic
  environments.  After $\approx 1$ Myr, these stars settle down to
  become steady sources of soft X-rays. I will use high-resolution
  X-ray spectroscopy and MHD modeling to show that young O stars like
  $\theta^1~{\rm Ori~C}$ are well explained by the magnetically
  channeled wind shock scenario.  After their magnetic fields
  dissipate, older O stars produce X-rays via shock heating in their
  unstable stellar winds.  Here too I will use X-ray spectroscopy and
  numerical modeling to confirm this scenario.  In addition to
  elucidating the nature and cause of the O star X-ray emission,
  modeling of the high-resolution X-ray spectra of O supergiants
  provides strong evidence that mass-loss rates of these O stars have
  been overestimated.

\end{abstract}

\firstsection 

\section{Introduction}

O stars dominate the X-ray emission from young clusters, with X-ray
luminosities up to $L_{\rm x} = 10^{34}$ ergs s$^{-1}$ and emission
that is hard (typically several keV) and often variable.  This strong
X-ray emission has an effect on these O stars' environments, including
nearby sites of star formation and protoplanetary disks surrounding
nearby low-mass pre-main-sequence stars.  The X-ray emission is also
interesting in its own right, as it traces important high-energy
processes in the extended atmospheres of O stars. In this paper, I
will show how the spectral properties of the X-rays from the
prototypical young, magnetized O star, $\theta^1~{\rm Ori~C}$ are in
line with the predictions of the Magnetically Channeled Wind Shock
(MCWS) model, but how this process seems to dissipate as O stars age,
with weaker line-driven instability wind shocks explaining the X-ray
emission in older O stars.  I also will show how the X-ray emission
can be used as a probe of the conditions in the bulk stellar winds of
these objects.  Specifically, the resolved X-ray line profiles in
normal O supergiants provide an independent line of evidence for
reduced mass-loss rates.

Now, it is certainly the case that many young O stars do not show the
X-ray signatures of the MCWS mechanism.  And indeed, only a handful of
O and early B stars have had direct detections of magnetic fields.  Of
course, highly structured, non-dipole fields will be very difficult to
detect on hot stars, even if their local strength is quite high.  But
one should keep in mind that wind-wind interactions in close binaries
can also produce the hard, strong, and variable X-rays seen in many
young O stars.

Because $\theta^1~{\rm Ori~C}$ has a well established, predominantly
dipole field, and because many of its properties are explained by this
field, I treat it here as a potential prototype.  I also note that the
incidence of hard, strong X-ray emission from O stars diminishes
rapidly as one looks from young ($<1$ Myr) clusters to older (2 to 5
Myr) clusters.  This fact can be explained if the fossil fields in
young O stars dissipate as the stars age. If wind-wind binaries
account for many of these sources, then only the very earliest O stars
with very short lifetimes are involved.

\section{Data comparison: high-resolution X-ray spectra}

\begin{figure}[h]
\begin{center}
 \includegraphics[scale=0.15]{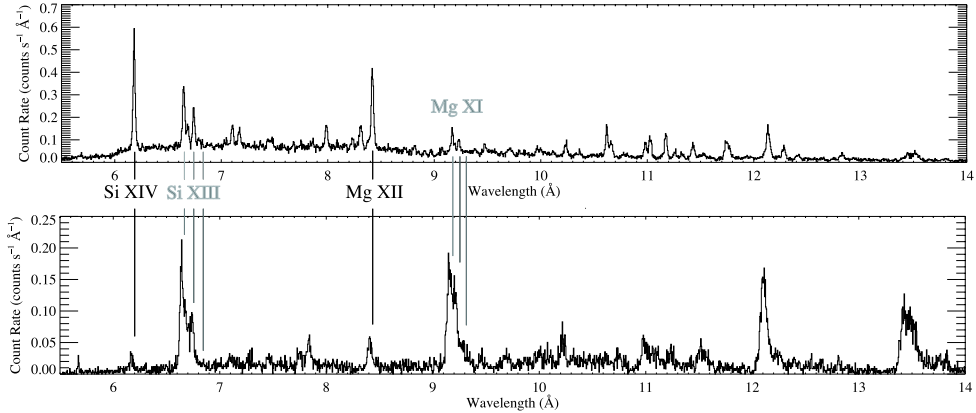} 
 \caption{{\it Chandra} MEG spectra of $\theta^1~{\rm Ori~C}$ (top)
   and $\zeta$ Pup (bottom). The hydrogen-like Lyman alpha lines of Si
   and Mg are indicated in black, while the helium-like
   resonance-intercombination-forbidden complexes of the same elements
   are indicated in gray. }
   \label{fig:spectra}
\end{center}
\end{figure}

We begin the comparison of young and old O stars and their X-rays by
showing in Fig.\ \ref{fig:spectra} the {\it Chandra} MEG spectra of
two representative stars: $\theta^1~{\rm Ori~C}$ (O4-7 V), with an age
of $\approx 1$ Myr, and $\zeta$ Pup (O4 If), with an age of several
Myr, and already evolved well off the main sequence. The $\zeta$ Pup
X-ray spectrum is typical of those measured for most O stars.

\begin{figure}[h]
\begin{center}
 \includegraphics[scale=0.23]{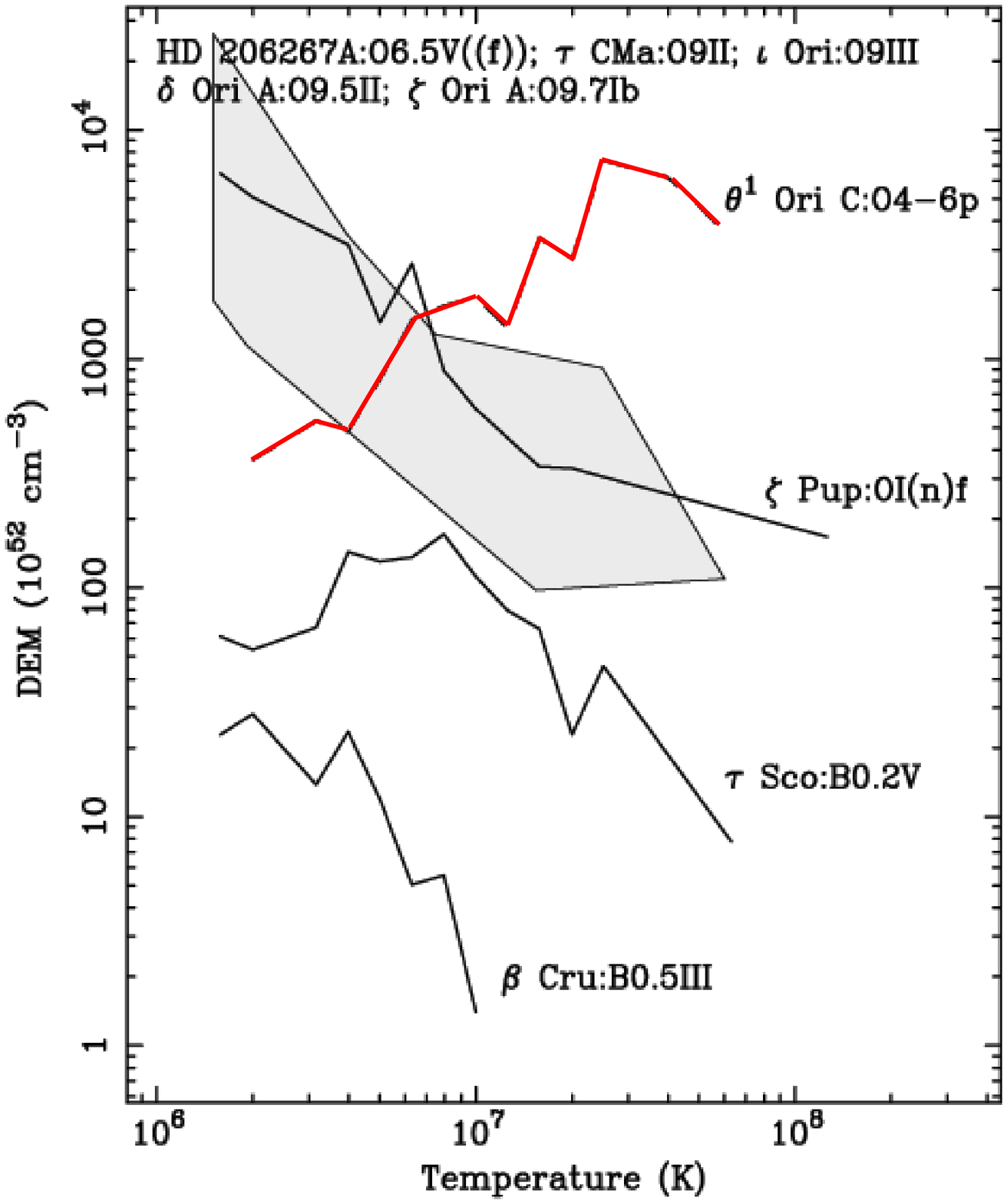} 
 \includegraphics[scale=0.30]{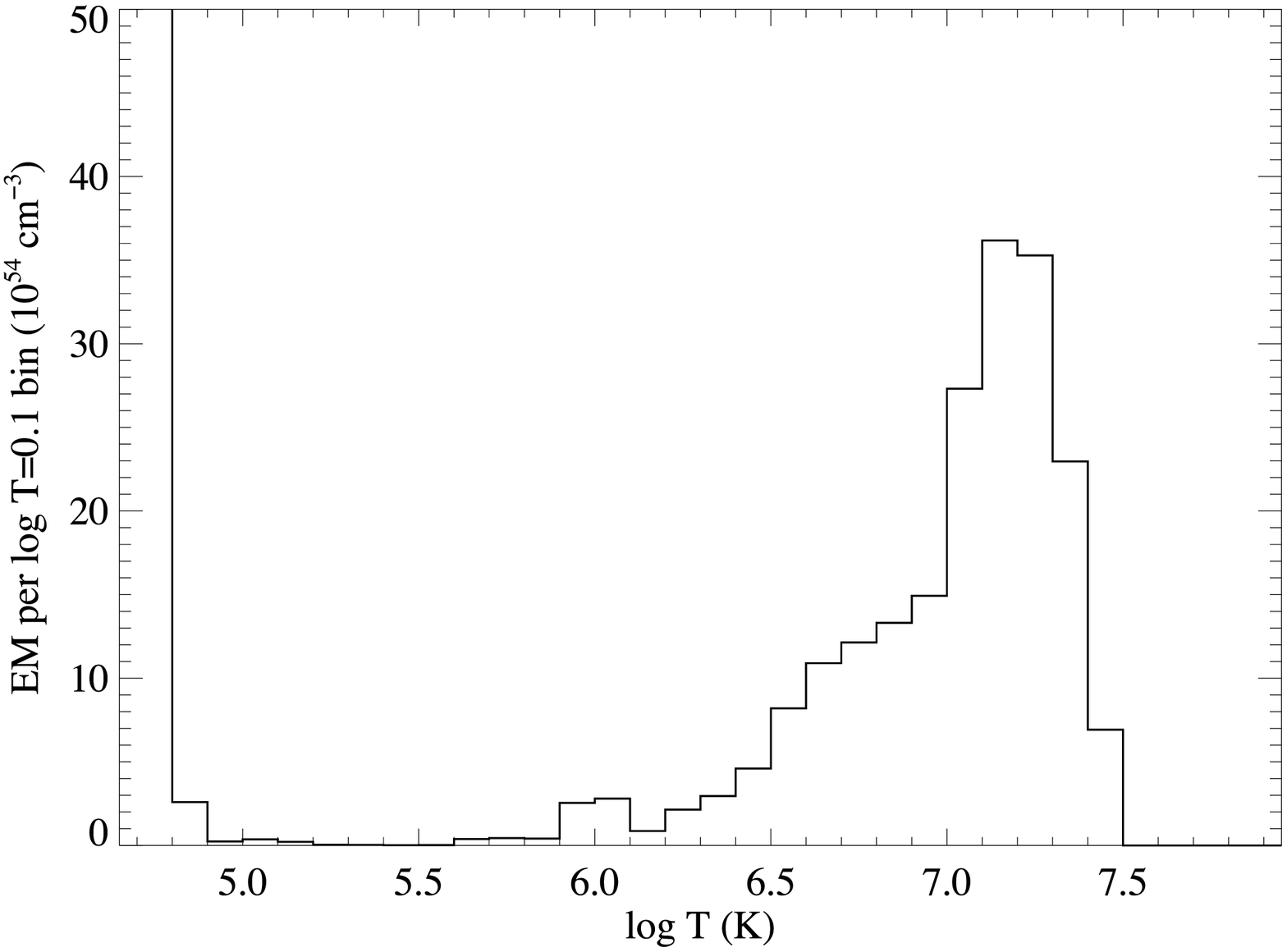} 
 \caption{Differential emission measures of seven O stars and two B
   stars, derived from thermal spectral model fits to {\it Chandra}
   spectra (left; taken from \cite[Wojdowski \& Schulz,
   2005]{ws2005}). $\theta^1~{\rm Ori~C}$ is the only star whose DEM
   has a positive slope. The panel on the right shows a DEM predicted
   by the MHD simulations of $\theta^1~{\rm Ori~C}$ (taken from
   \cite[Gagn\'{e} et al., 2005]{Gagne2005}). }
   \label{fig:dem}
\end{center}
\end{figure}

Two obvious differences between these spectra are the hardness of the
X-ray emission from $\theta^1~{\rm Ori~C}$ and the small line widths
in that star's spectrum.  The hardness implies a much higher plasma
temperature in the young O star, and this is best seen in the data
when one compares resonance lines of hydrogen-like and helium-like
ionization states of abundant elements.  In Fig.\ \ref{fig:spectra} I
have labeled these lines for silicon and magnesium.  In $\theta^1~{\rm
  Ori~C}$ the hydrogen-like lines are much stronger, whereas in
$\zeta$ Pup the helium-like lines are stronger.  This reflects a
significantly different ionization balance in these two stars which is
a direct effect of their different plasma temperatures.  The plasma
temperature distribution, based on the analysis of high-resolution
{\it Chandra} spectra of several O stars, has been determined by
Wojdowski \& Schulz (2005).  These authors assume a continuous
Differential Emission Measure (DEM), which can be thought of as a
density-squared weighting of the plasma temperature distribution.
Their results are shown in Fig.\ \ref{fig:dem}, where it can easily be
seen that the DEM for $\theta^1~{\rm Ori~C}$ is the only one (of seven
O stars) with a positive slope.  Its peak is near $T=30$ million K,
whereas the DEMs of the more evolved O stars (including $\zeta$ Pup)
all peak near one or two million K.

\begin{figure}[h]
\begin{center}
 \includegraphics[scale=0.38,angle=0]{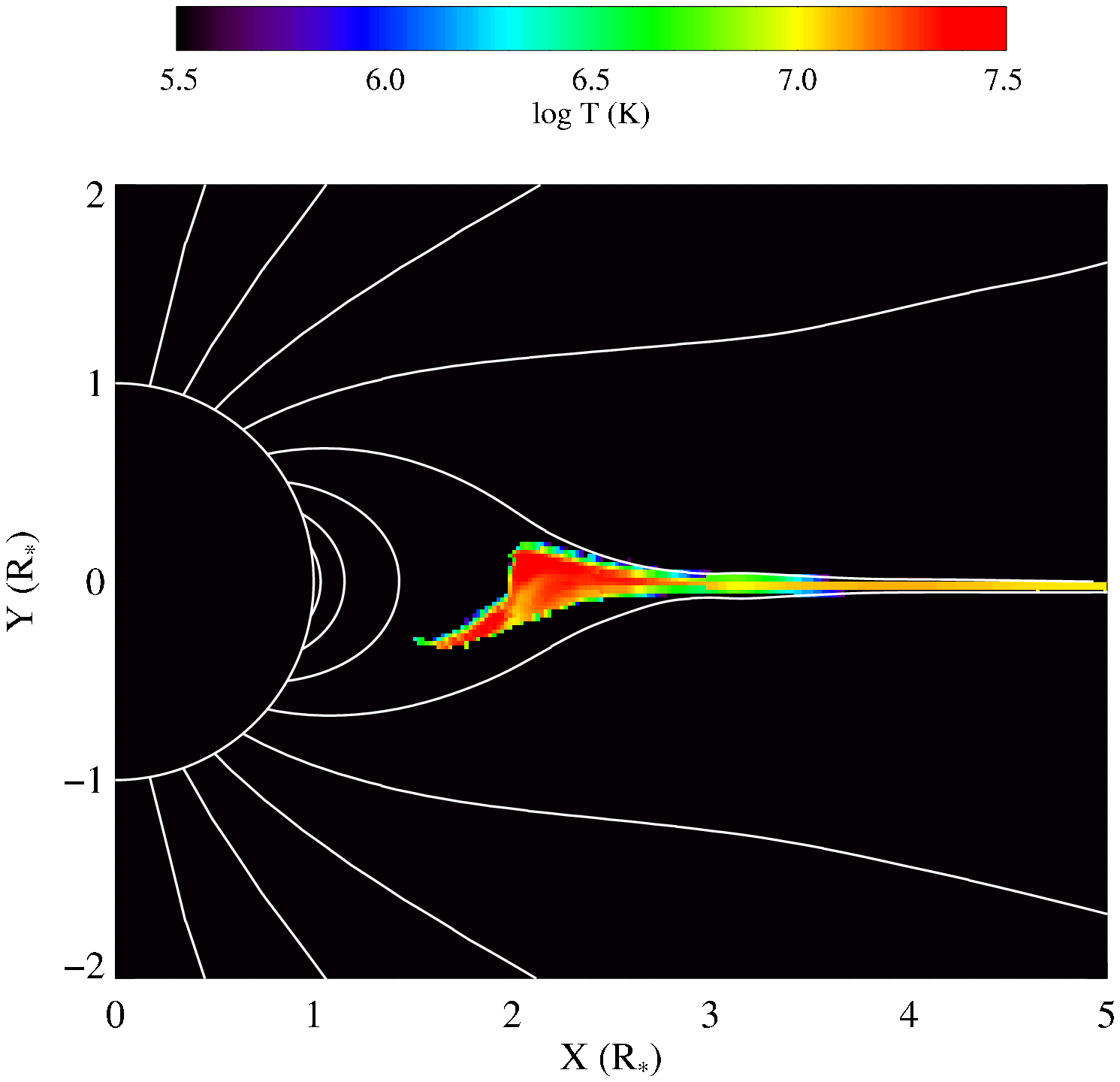} 
 \includegraphics[scale=0.38,angle=0]{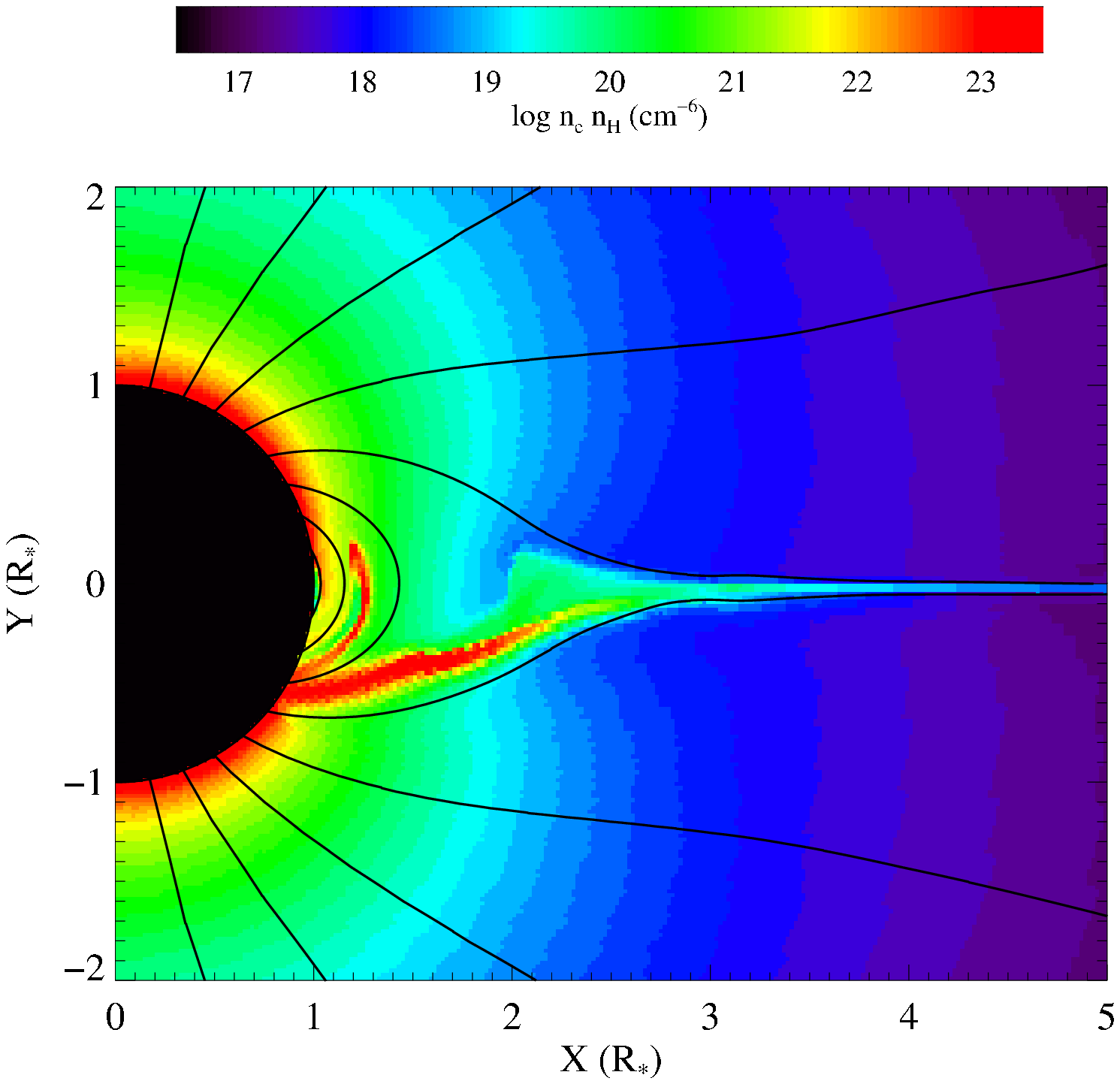} 
 \caption{Contour plots of temperature (left) and emission measure per
   unit volume (right) from a 2-D MHD simulation of the magnetically
   confined wind of $\theta^1~{\rm Ori~C}$ (taken from \cite[Gagn\'{e}
   et al., 2005]{Gagne2005}).  Magnetic field lines are displayed as
   contours.  The wind flow up each closed field line encounters a
   strong shock due to the ram pressure of the wind flow from the
   opposite hemisphere, which heats the plasma according to
   $T_{\mathrm {shock}} \approx 10^7(v_{\mathrm {shock}}/1000~{\mathrm
     {km~ s^{-1}}})^2$ K. The head-on nature of the wind shocks leads
   to high shock velocities and temperatures.  In these MHD models,
   the field configuration is self-consistently solved for along with
   the wind dynamics. Note that another difference between the MHD
   simulations of the MCWS model and the initial analysis of
   \cite[Babel \& Montmerle (1997)]{bm1997} is the dynamical infall of
   material from the magnetic equator. Evidence of this can be seen in
   the snake-like structure visible in the emission measure panel,
   just above the star's surface, slightly below the equator. }
   \label{fig:mhd}
\end{center}
\end{figure}

We also show in Fig.\ \ref{fig:dem} the DEM from a snapshot of an MHD
simulation of the magnetically confined wind of $\theta^1~{\rm
  Ori~C}$.  The agreement is quite good, both in terms of the overall
emission measure and the shape of the DEM.  The simulation shows a
peak at 15 to 20 million K, modestly lower than that seen in the data,
but the main property -- a rising DEM up to and beyond 10 million K --
matches the data well. These MHD simulations confirm the predictions
of \cite{bm1997} that strong shock fronts near the magnetic equator
where the wind from the northern and southern hemispheres meets can
heat a significant amount of wind plasma to the observed high
temperatures.  Snapshots of temperature and emission measure from an
MHD simulation are shown in Fig.\ \ref{fig:mhd}. It can be seen that
the bulk of the shock-heated plasma is in the magnetically confined
region near $r \approx 2~{\mathrm {R_{\ast}}}$.  Due to this
confinement, the speed and line-of-sight velocity of this material is
low, and thus the emission lines from the thermal X-ray emission are
relatively narrow.

\begin{figure}[h]
\begin{center}
 \includegraphics[scale=0.27,angle=90]{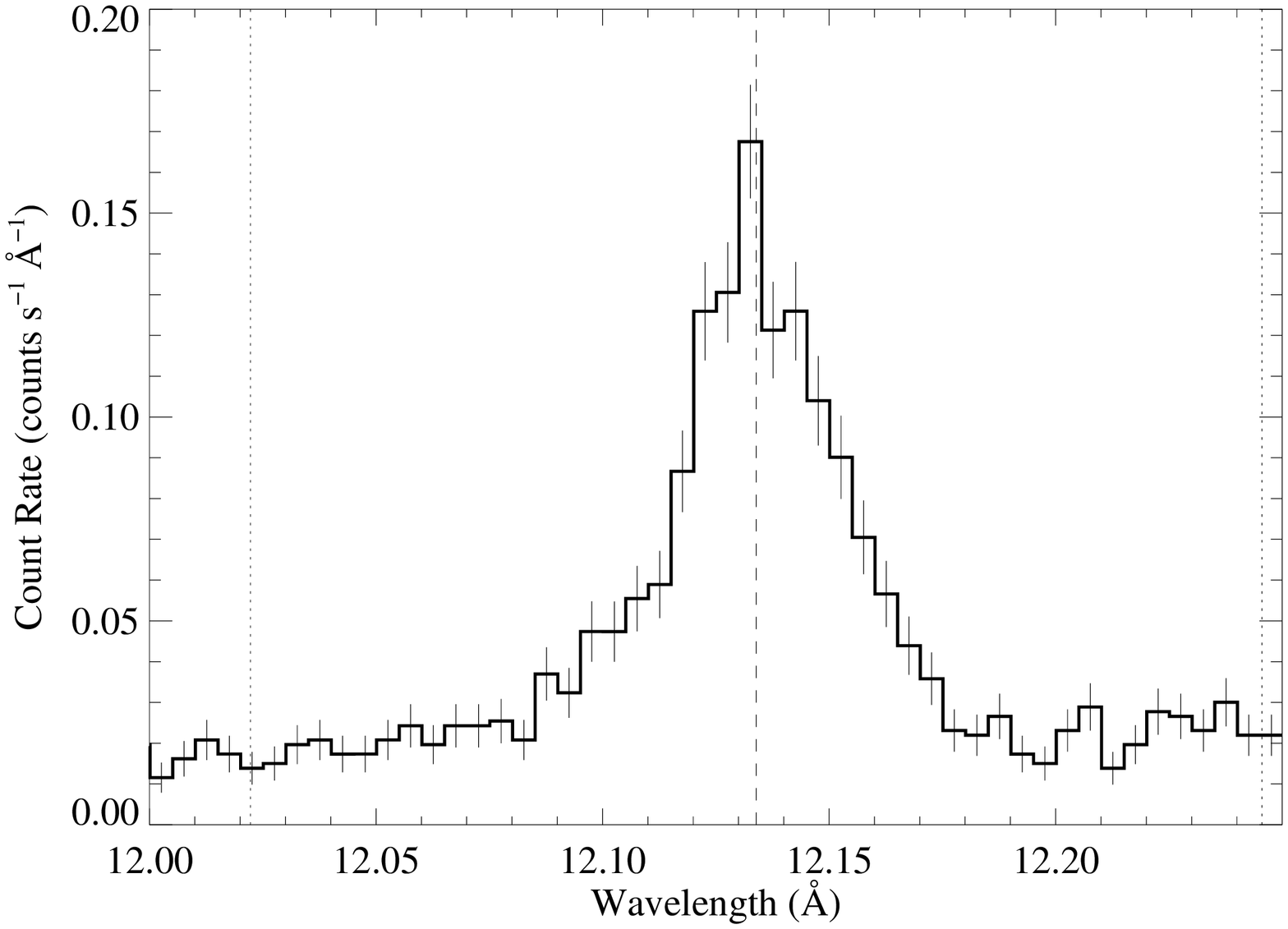} 
 \includegraphics[scale=0.27,angle=90]{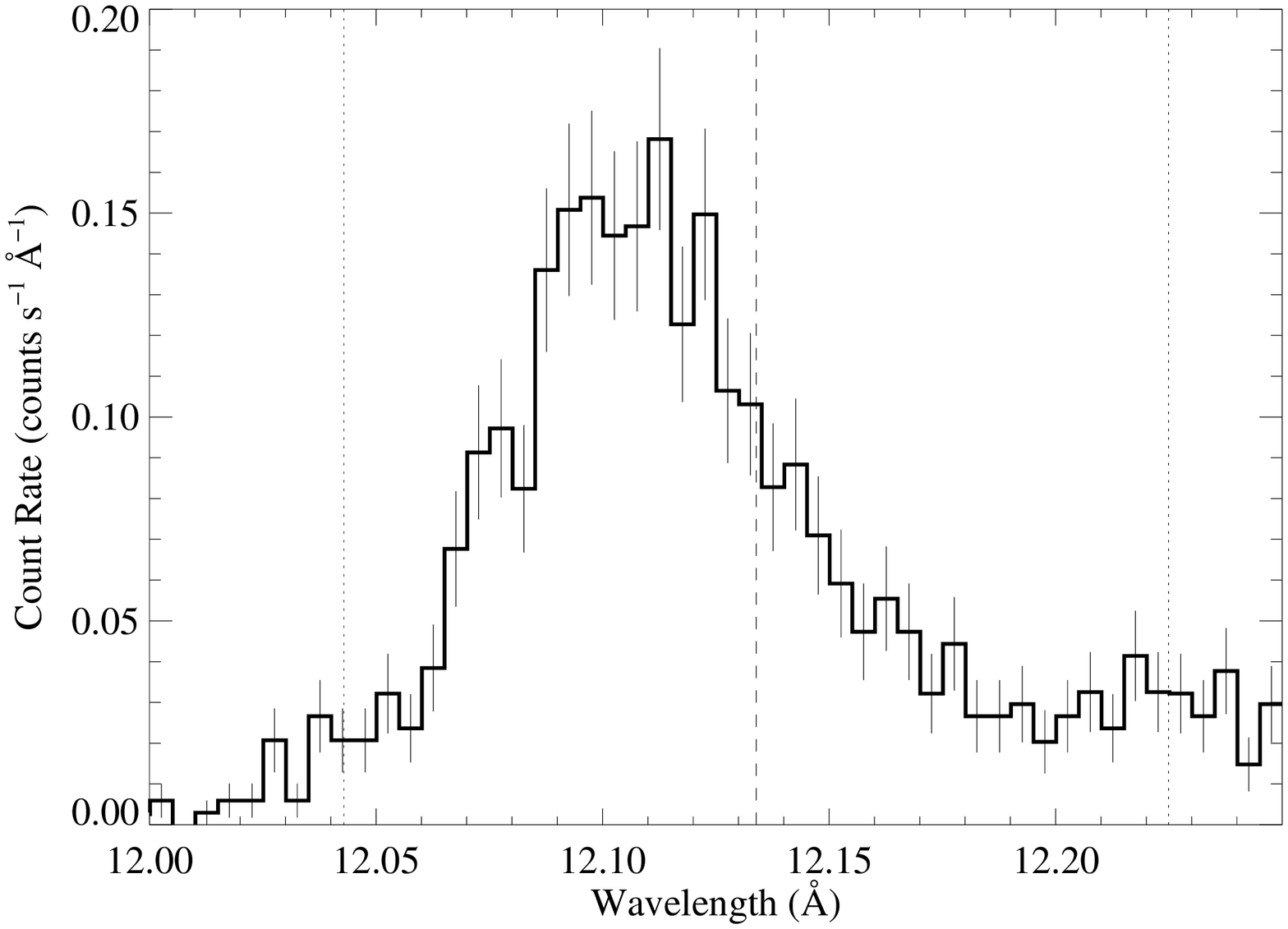} 
 \caption{Ne X Lyman-alpha lines in the {\it Chandra} MEG spectra of
   $\theta^1~{\rm Ori~C}$ (left) and $\zeta$ Pup (right). The vertical
   dashed line represents the laboratory rest wavelength of this
   transition, and the vertical dotted lines represent the blue and
   red shifts associated with the UV wind terminal velocity in each
   star.  Note that the profile in the $\zeta$ Pup spectrum is shifted
   and skewed as well as being broadened. Error bars are from Poisson
   photon-counting statistics. }
   \label{fig:line_profiles}
\end{center}
\end{figure}

\begin{figure}[h]
\begin{center}
 \includegraphics[scale=0.7,angle=0]{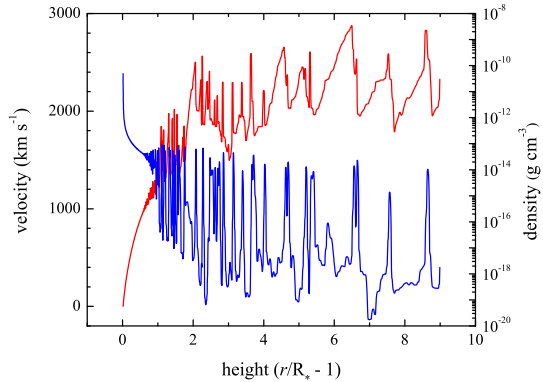} 
 \caption{A snapshot showing the velocity (red; left-hand
   axis) and density (blue; right-hand axis) as a function of
   height above the photosphere in a radiation hydrodynamics
   simulation of the wind of $\zeta$ Pup.  Shock-heated plasma cools
   rapidly; the bulk of the wind in any given snapshot is cold. }
   \label{fig:vh1}
\end{center}
\end{figure}

In contrast, the X-ray emission lines of mature O stars, like $\zeta$
Pup, are quite broad, as can be seen in Fig.\ \ref{fig:line_profiles},
where I show the neon Ly$\alpha$ lines for the two stars. The X-ray
spectrum of $\zeta$ Pup is also soft, as I have already shown. The
X-ray emission from these older, presumably non-magnetized, O stars is
thought to arise in much milder wind shocks, embedded in the
outflowing, highly supersonic line-driven winds.  The Line-Driven
Instability (LDI) is generally thought to produce these shocks,
although models have difficulty reproducing the overall level of X-ray
emission unless the instability is seeded, perhaps by sound waves
injected at the base of the wind (\cite[Feldmeier et al.,
1997]{fpp1997}).  The softness of the X-ray spectra, along with the
large line widths from the high velocity of the shock-heated wind, is
well explained by this LDI wind-shock scenario, as I show in Fig.\
\ref{fig:vh1}. This figure shows a snapshot from a 1-D radiation
hydrodynamics simulation of the wind of an O supergiant like $\zeta$
Pup, accounting for non-local line radiation transport.  The
instability grows rapidly beyond about half a stellar radius (in
height; $r=1.5~{\mathrm {R_{\ast}}}$).  Shock fronts can be seen in
this snapshot, but they typically have velocities of only a few
hundred km s$^{-1}$, leading to heating of only a few million K.  The
soft spectra seen in Fig.\ \ref{fig:spectra} and the DEMs weighted to
low plasma temperatures, shown in the left-hand panel of Fig.\
\ref{fig:dem}, are in line with the results of this simulation.

\begin{figure}[h]
\begin{center}
 \includegraphics[scale=0.5,angle=0]{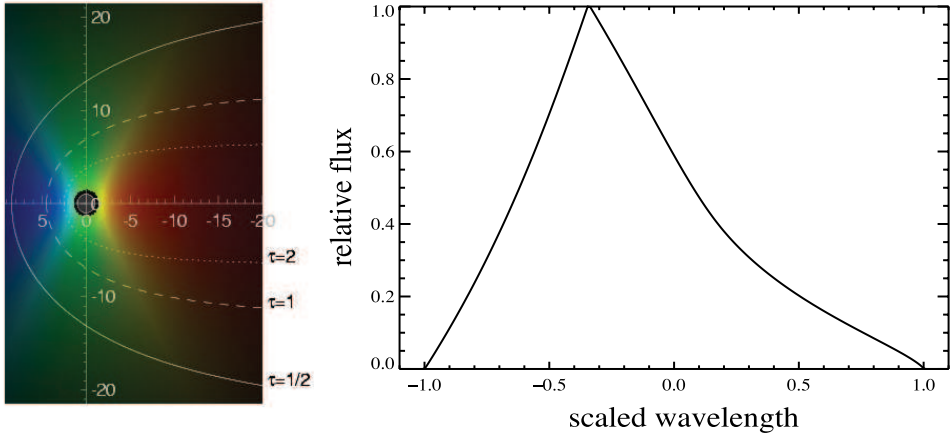} 
 \caption{Schematic representation (left) of the spherically symmetric
   stellar wind model underlying the simple, empirical X-ray emission
   line profile models that I fit to the data.  Here the emission
   measure of the hot, emitting component of the wind is indicated
   according to brightness, while its line-of-sight velocity with
   respect to an observer on the left is indicated by color, from blue
   on the front side to red on the back.  The units of the axis labels
   are stellar radii.  Note that the emission starts half a stellar
   radius above the surface of the star (which can be seen as a white
   circular outline), and its intensity falls off with radius.  The
   absorption properties of the dominant, cool wind component in this
   model are represented by the three white contours of constant
   optical depth (again, for an observer on the left).  The emission
   line profile that results from this particular model realization is
   shown (at infinite resolution) in the right-hand panel.  Negative
   values of the scaled wavelength correspond to blue shifts. The
   effect of the continuum opacity on the skewed line profile can be
   clearly seen.  Note that most of the red-shifted, back side of the
   stellar wind suffers at least one optical depth of attenuation. }
   \label{fig:profile_schematic}
\end{center}
\end{figure}

The predictions of this model can be further tested by examining the
X-ray emission line profiles, which have a characteristic asymmetric,
skewed shape, as can be seen in the right-hand panel of Fig.\
\ref{fig:line_profiles}. This shape, with a deficit of red-shifted
photons, arises in a spherically expanding wind with hot,
line-emitting material intermixed with warm, continuum-absorbing
material, as motivated by the global structure seen in Fig.\
\ref{fig:vh1}.  If the wind is optically thick (in the continuum;
$\kappa \approx$ constant across a line), then there is significantly
more attenuation of emission from the back of the wind, which is the
red shifted portion.  And there is comparatively more emission from
the front, blue shifted, less attenuated, side. In Fig.\
\ref{fig:profile_schematic} I show a schematic demonstrating a simple,
empirical wind line profile model that I use to fit the data.

\begin{figure}[h]
\begin{center}
 \includegraphics[scale=0.4,angle=90]{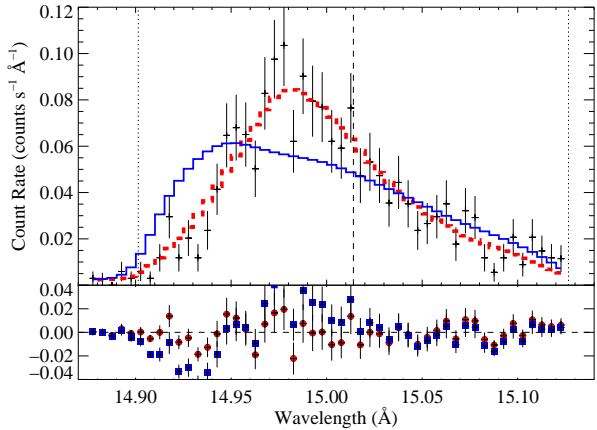} 
 \caption{Best-fit wind profile model, for a non-porous wind, (red
   dashed histogram) fit to the Fe\, {\sc xvii} line in the {\it
     Chandra} spectrum of $\zeta$ Pup. The (blue) solid histogram is
   the best-fit non-porous model for which the optical depth parameter
   is fixed at the value implied by the literature mass-loss rate
   ($\tau_{\mathrm {\ast}}=8$). Fit residuals for the two models are
   shown in the lower panel, with the red circles corresponding to the
   low optical depth model represented by the dashed histogram, and
   the blue squares corresponding to the higher optical depth model.}
   \label{fig:2fits}
\end{center}
\end{figure}

I fit an empirical model (\cite[Owocki \& Cohen, 2001]{oc2001}) to
emission lines in the spectrum of $\zeta$ Pup and get good fits by
adjusting only three parameters: the normalization, the inner radius
below which there is assumed to be no emission ($R_{\rm o}$), and the
wind optical depth (parameterized by the quantity $\tau_{\rm \ast}
\equiv {\dot{M} \kappa} / {4 \pi R_{\rm \ast} v_{\rm \infty}}$). The
fit to the Fe\, {\sc xvii} line at 15.014 \AA\ is typical and is shown
in Fig.\ \ref{fig:2fits} as the red dashed histogram. The fit is
formally good, and the best-fit values with joint 68\% parameter
confidence limits are $R_{\rm o} = 1.53^{+0.12}_{-0.15} ~{\rm
  R_{\ast}}$ and $\tau_{\rm \ast} = 2.0 \pm 0.4$.

\section{Evidence for a lowered mass-loss rate}

The $R_{\rm o}$ derived from the profile is consistent with the onset
of the self-excited instability seen in the hydrodynamics simulation
shown in Fig.\ \ref{fig:vh1}. The optical depth consistent with the
data is actually quite small, which can be seen qualitatively in the
relatively modest asymmetry.  The mass-loss rate of $6 \times 10^{-6}~
{\mathrm {M_{\odot}~yr^{-1}}}$ derived from H$\alpha$ emission
(\cite[Puls et al., 1996]{Puls1996}) implies $\tau_{\rm \ast} = 8$.
The best-fit model with that value fixed is shown in Fig.\
\ref{fig:2fits} as the blue solid histogram. The fit is formally very
poor.  Thus it would appear that the X-ray line profiles provide
independent evidence that mass-loss rates of O stars must be revised
downward by a factor of several; a factor of 4 for this star,
according to this particular line ($\tau_{\mathrm {\ast}}=2.0 \pm 0.4$
vs.\ $\tau_{\mathrm {\ast}}=8$).

\begin{figure}[h]
\begin{center}
 \includegraphics[scale=0.315,angle=90]{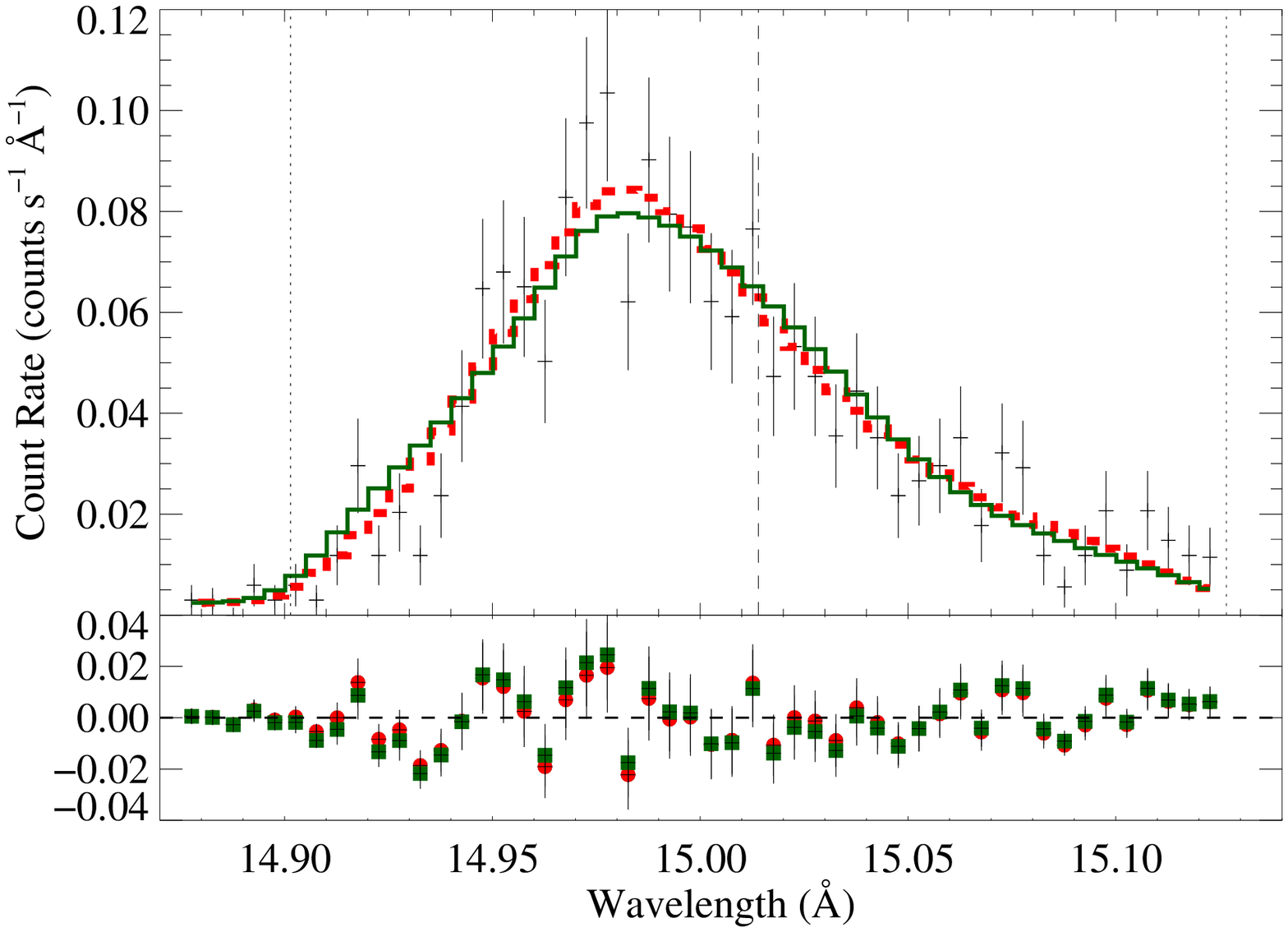} 
 \includegraphics[scale=0.31,angle=0]{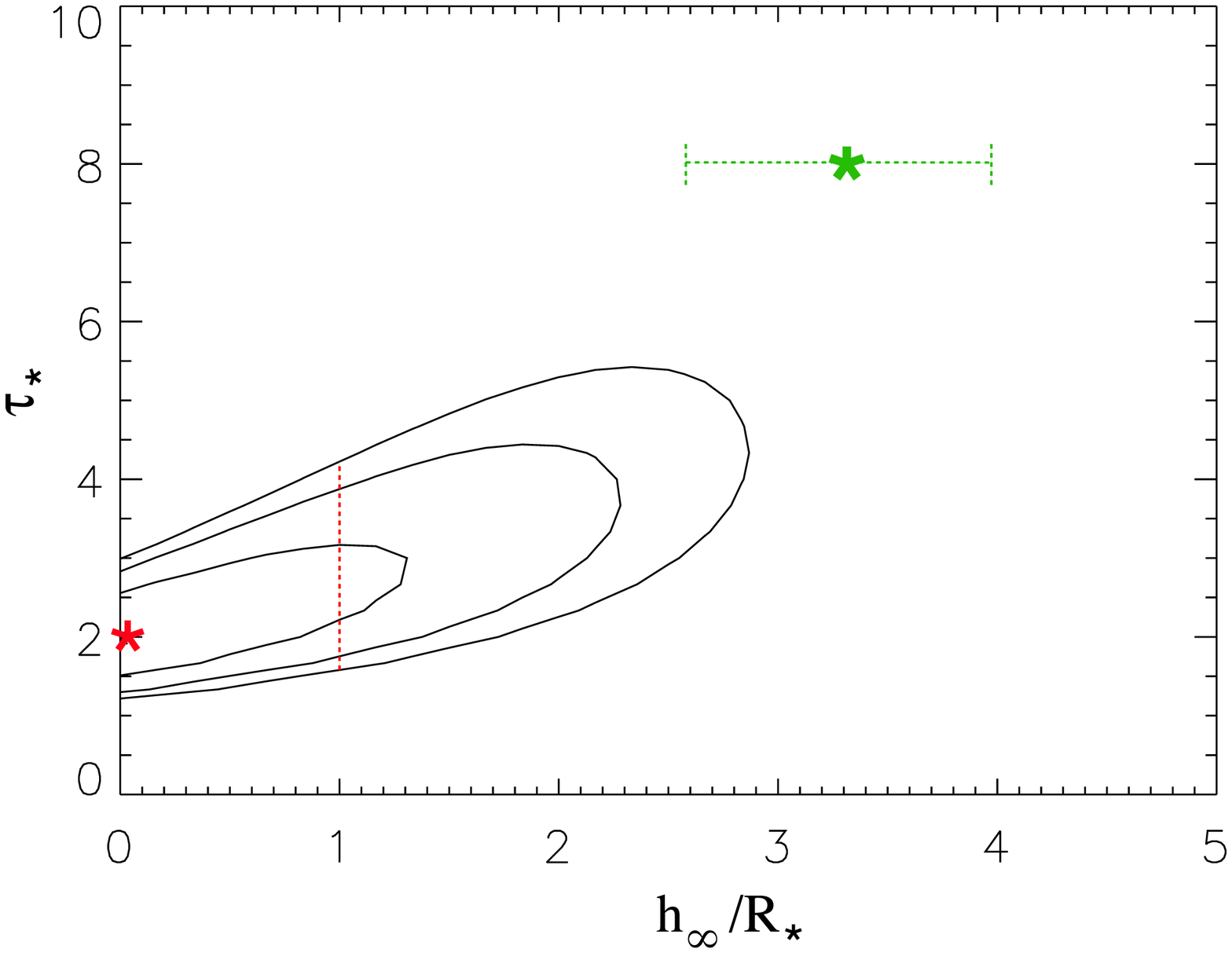} 
 \caption{Best-fit wind profile model, for a non-porous wind (red
   dashed histogram), fit to the Fe\, {\sc xvii} line in the {\it
     Chandra} spectrum of $\zeta$ Pup (left). The green solid
   histogram is the best-fit porous model for which the optical depth
   parameter is fixed at the value implied by the literature mass-loss
   rate ($\tau_{\rm \ast} = 8$). We also show confidence limits (68,
   90, 95\%) in $h_{\mathrm {\infty}} {\rm {vs.}}~\tau_{\mathrm
     {\ast}}$ parameter space (right). The global best-fit model is
   indicated by the red star at $h_{\mathrm {\infty}} = 0,
   \tau_{\mathrm {\ast}} = 2$.  The green star at $h_{\mathrm
     {\infty}} = 3.3, \tau_{\mathrm {\ast}} = 8$ represents the
   best-fit model with the wind optical depth fixed at the value
   implied by the literature mass-loss rate.  The green horizontal bar
   centered on it represents the 68\% confidence limit on the value of
   $h_{\mathrm {\infty}}$, given $\tau_{\mathrm {\ast}} = 8$ ($2.5 <
   h_{\mathrm {\infty}}/{\mathrm R_{\ast}} < 4.0$).  The red vertical
   line at $h_{\mathrm {\ast}} = 1$ emphasizes that even that large
   porosity length cannot bring the optical depth close to the value
   associated with the literature mass-loss rate. }
   \label{fig:2fits_porous}
\end{center}
\end{figure}

It has been suggested that porosity associated with large-scale
clumping -- rather than reduced mass-loss rates -- can account for the
surprisingly small degree of asymmetry in the observed X-ray emission
line profiles in O stars (\cite[Oskinova et al., 2006]{ofh2006}).  By
fitting the line profile model, modified for the effects of porosity
produced by spherical clumps (\cite[Owocki \& Cohen, 2006]{oc2006}), I
can quantify the trade-off between atomic opacity and porosity (see
also \cite[Cohen et al., 2008]{Cohen2008}).  In Fig.\
\ref{fig:2fits_porous} I show the best-fit porous model with
$\tau_{\rm \ast} = 8$, the value implied by the literature mass-loss
rate, and with the porosity length, $h \equiv \ell/f$, free to vary.
Here $\ell$ is the clump size and $f$ is the volume filling factor of
the clumps.  The porosity length, $h$, completely describes the
effects of porosity on line profiles and in the limit of small clumps
it is equivalent to the interclump spacing.  The best-fit porous model
with the literature mass-loss rate is nearly indistinguishable from
the best-fit non-porous model, although the fit quality is formally
not as good.  More importantly, it requires a terminal porosity length
(the value of $h$ in the outer portion of the wind) of at least
$h_{\mathrm {\infty}} = 2.5$ R$_{\rm \ast}$ (68\% confidence), as
shown in the right-hand panel of Fig.\ \ref{fig:2fits_porous}. Even
ignoring the worse quality of the porous model fits, the very high
values of the porosity length required to fit the data are vastly
larger than the porosity lengths seen in state-of-the-art 2-D
radiation hydrodynamics simulations, which show LDI-induced structure
down to the grid scale (\cite[Dessart \& Owocki,
2005]{do2005}). Therefore, I conclude that there is no compelling
evidence that porosity explains the modestly asymmetric X-ray
line profiles in $\zeta$ Pup. Rather, these profiles provide
independent evidence for a reduced mass-loss rate.

\noindent
I acknowledge support from NASA/CXC grant AR7-8002X and from the
Hungerford Faculty Support Endowment of Swarthmore College's provost's
office.

\begin{discussion}

  \discuss{Kudritzki}{It is very important to repeat the observing
    experiment that you have done for theta-1 Ori C for young O stars,
    which are more massive and luminous than theta-1 Ori C.  I
    speculate that for those objects, the magnetic focusing of the
    stellar winds will be less effective, because the winds are
    stronger and the ratio of magnetic to mechanical wind energy is
    lower.  I think it is really crucial to do such observations.}

  \discuss{Cohen}{I agree.  The degree of channeling and confinement,
    however, goes as B$^2$ but only as 1/M-dot, so the extent of
    channeling and confinement in any given star is more likely to be
    dominated by trends in magnetic field strength, which we don't
    understand, than by trends in mass-loss rate associated with
    stellar mass and luminosity.  I think the key measurements to make
    in order to test the idea that the MCWS mechanism on theta1 Ori C
    is a paradigm for X-ray emission in young O stars would be Zeeman
    measurements of fields on the massive cluster stars that are known
    to be strong, hard X-ray sources.  High-resolution X-ray
    spectroscopy would also be very useful, obviously, if it's
    feasible.  As I discussed in my talk in the special session on
    magnetic massive stars on Sunday, line widths and helium-like
    forbidden-to-intercombination line strength ratios provide
    information beyond what's provided by CCD-based (e.g.\ Chandra
    ACIS or XMM EPIC) X-ray data.}

  \discuss{Zinnecker}{My question refers to theta-1 Ori C and the
    origin of the obliquity between its rotational axis and magnetic
    field axis.  From star formation theory it would seem an {\it
      aligned} magnetic rotator would be expected.  Any suggestions to
    explain the misalignment?}

  \discuss{Cohen}{I don't have any special expertise in star formation
    theory, but if models predict aligned rotation and magnetic axes,
    then there must be some important physics missing from them. Many
    of the magnetic massive stars have highly misaligned axes: tau
    Sco, beta Cep, and sigma Ori E, for example, are all close to
    $\beta = 90$ degrees. }

  \discuss{Skinner}{ A new paper appearing in the journal {\it
      Science} this week by M. G\"{u}del et al. (2007) reports the
    first detections of hot (1 to 2 MK) {\it diffuse} X-ray emission
    in the extended Orion Nebula.  This article argues that massive
    Trapezium O stars (and their shocked winds) are ultimately
    responsible for the diffuse X-ray emission detected by
    XMM-Newton.}

\discuss{Cohen}{I think that the problem of heating the diffuse, X-ray
  emitting plasma in massive star clusters is a hard one.  The shocked
  wind (such as zeta Pup's) will adiabatically cool over distances
  much less than 1 pc. And the morphology of the X-rays isn't what
  you'd expect from the wind slamming into dense interstellar gas at
  the boundaries of these cavities.}

\discuss{Walborn}{ There are currently only two O stars with observed
  magnetic fields, theta-1 Ori C and HD 191612.  The latter can be
  understood as a spun-down version of the former, with a rotational
  period of 538 d (vs. 15 d for theta-1 Ori C), a soft X-ray spectrum,
  and an age of 3 to 4 Myr.  Nevertheless, it is a very unusual
  object, with a very peculiar spectrum and extreme, periodic spectral
  variations.  So these two stars are inconsistent with your
  hypothesis that theta-1 Ori C is a typical very young O star, and
  that magnetic fields have disappeared at $\sim 5$ Myr.  I think that
  both of these objects are unusual areas of large fossil fields.  An
  alternate interpretation of your cluster X-ray differences might be
  different frequencies of wind-wind collision binaries.}

\discuss{Cohen}{While I agree than wind-wind collision binaries may
  make a significant contribution to the observed population of hard,
  strong X-ray sources in young clusters, I think we need more
  information about the highly unusual O star, HD 191612, not to
  mention more positive detections or strong upper limits to magnetic
  field strengths in other O stars, both young and old.  The X-ray
  spectrum of HD 191612 looks like that of a typical, older O star,
  with very broad lines and an SED that is quite soft.  Perhaps the
  fields of young O stars like theta-1 Ori C become more spatially
  structured as they evolve.  It's possible that HD 191612 has a field
  that looks more like that of tau Sco and is not dominated by a large
  scale dipole.  If that's the case, then there may not be any
  large-scale confinement and channeling of the stellar wind, and no
  substantial MCWS mechanism.  The X-ray emission may instead arise in
  open field regions, in a loose analogy to the solar wind and coronal
  holes. }

\end{discussion}

\end{document}